\documentclass[11pt]{article}
\usepackage[latin1]{inputenc}
\usepackage[dvips]{epsfig}
\usepackage{amsfonts}
\newcommand{\ew}{\mbox{$\epsilon$}}

\newcommand{\nrep}[3]{#1^{#2..#3}}
\newcommand{\RENUM}{\mbox{\#RE}}

\newtheorem{Theorem}{Theorem}[section]
\newcommand{\Proof}{\textbf{Proof}}
\newcommand{\Done}{\hspace{1pt} \hspace*{\fill} $\Box$} 
\newcommand{\seq}[3]{#1_{#2}, \ldots , #1_{#3}}
\newcommand{\upto}[2]{\seq{#1}{1}{#2}}

\newcommand{\OR}{\mbox{$\mid$}}

\begin{document}
\title{
Inclusion of Unambiguous \RENUM{}s
is NP-Hard
}
\author{Pekka Kilpeläinen 
\\
University of Kuopio 
\\
Department of Computer Science 
\\
{\tt Pekka.Kilpelainen@cs.uku.fi }
}
\date{May 27, 2004}
\maketitle
\begin{abstract}
We show that testing  inclusion between languages represented
by 
regular expressions with numerical occurrence indicators (\RENUM{}s)
is NP-hard, even if the expressions satisfy the requirement of
``unambiguity'', which is required for 
XML Schema content model expressions. 
\end{abstract}

\section{Proof of the result}
We have seen before~\cite{KilTuh03} that 
testing for inclusion and overlap of languages represented by
\RENUM{}s is NP-hard. Testing for the overlap was seen hard
also for expressions that satisfy the XML requirement of
``unambiguity''. On the other hand, the NP-hardness proof of
\RENUM\ inclusion used ambiguous expressions.
Here we show that unambiguity does not make the testing 
of inclusion essentially easier.
The proof is based on a polynomial time
Turing reduction~\cite[Chap.~5]{GareyJohnson79} from PARTITION, which is one of the
best-known NP-complete problems~\cite{Karp72,GareyJohnson79}.

\begin{Theorem}
The \RENUM\ inclusion problem is NP-hard, also for unambiguous
\RENUM{}s.
\end{Theorem}
\Proof. 
Let a set $A = \{\upto{a}{k}\}$ and a positive integer weight $w(a)$
of each $a \in A$ form an instance of PARTITION.
The problem is to decide whether $A$ can be split in two
equal-weight subsets $A'$ and $A-A'$, that is, whether 
\begin{equation}
\label{eq:PART}
\sum_{a \in A'} w(a)\ = \sum_{a \in A-A'} w(a)
\end{equation}
holds for some $A' \subseteq A$.
Notice that (\ref{eq:PART}) can hold only if the total weight
of the set $A$ is even. Therefore we can assume that 
$
\sum_{a \in A} w(a) = 2 n 
$ 
for some positive integer $n$, 
which means that (\ref{eq:PART}) holds if and only if 
\begin{equation}
\label{eq:PART2}
\sum_{a \in A'} w(a) = n
\end{equation}
for some $A' \subseteq A$.

For shortness, denote the weight $w(a_i)$ of an item $a_i \in
A$ by $w_i$.

Now form the following two \RENUM{}s over the alphabet 
$\Sigma = \{a_0, \upto{a}{k} \}$:
\begin{eqnarray*}
E_1 &=& \nrep{a_0}{n+1}{n+1}(\nrep{a_1}{w_1}{w_1} \OR
\ew)(\nrep{a_2}{w_2}{w_2} \OR \ew) \cdots
(\nrep{a_k}{w_k}{w_k} \OR \ew)
\\
E_2 &=& \nrep{
(\nrep{(a_0 \OR a_1 \OR \cdots \OR a_k)}{n+1}{2n})}{1}{2}
\end{eqnarray*}
Notice that both expressions are trivially unambiguous
since each symbol of $\Sigma$ appears exactly once in both of them. 
Expression $E_1$ describes words 
of the form $a_0^{n+1}u$, where the length of the suffix $u$ 
equals the total weight of some subset of $A$.
Therefore $L(E_1) \subseteq \{v \in \Sigma^* \mid n+1 \leq |v|
\leq 3n+1 \}$.
Obviously $E_1$ accepts a word of
length $2n+1$ if and only if a partition that satisfies
(\ref{eq:PART2}) exists. 
Expression $E_2$, on the other hand, \emph{rejects} 
any words of length $2n+1$:
\begin{eqnarray*}
L(E_2) &=& \bigcup_{i=n+1}^{2n}\Sigma^i\ \cup 
\bigcup_{i=2n+2}^{4n}\Sigma^i
\\
	&=&  \{v \in \Sigma^* \mid  n+1 \leq |v|
\leq 4n, |v| \neq 2n+1\}
\end{eqnarray*}
Now $L(E_1) \subseteq L(E_2)$ holds iff $E_1$ does not accept
any word of length $2n+1$, which holds if and only if
no partition which satisfies
(\ref{eq:PART}) exists.
\Done

So, a polynomial-time algorithm for testing the inclusion of
unambiguous \RENUM{}s would imply $P = NP$, which is
considered most unlikely.

\end{document}